\documentstyle[aps,twocolumn,epsf]{revtex}
\begin{document}
\draft
\title{Gravitational waves from quasi-spherical black holes}
\author{Sean A. Hayward}
\address{Center for Gravitational Physics and Geometry,
104 Davey Laboratory, The Pennsylvania State University,
University Park, PA 16802-6300, U.S.A.\\
{\tt hayward@gravity.phys.psu.edu}}
\date{21st January 2000}
\maketitle
\begin{abstract}
A quasi-spherical approximation scheme,
intended to apply to coalescing black holes,
allows the waveforms of gravitational radiation to be computed 
by integrating ordinary differential equations.
\end{abstract}

The coalescence of binary black holes is expected to be 
one of the main astrophysical sources for upcoming gravitational-wave detectors.
The initial phase of inspiral and the final phase of ringdown are understood
in terms of post-Newtonian and close-limit approximations respectively,
but the coalescence is only qualitatively understood and generally thought 
to be tractable only by numerical methods\cite{BBH1,BCT,P,BBH2}.
Considerable problems have been encountered 
and currently there are no reliable predictions of waveforms.

This article presents an approximation scheme 
to compute the gravitational waveforms 
for space-times close to spherical symmetry.
This is intended to apply to binary black holes once they have coalesced,
i.e.\ when a marginal surface encloses both sources.
The quasi-spherical approximation will be best when 
the angular momentum is small,
but note that 
even the maximally rotating Kerr black hole is $70\%$ spherically symmetric 
according to the ratio of the areal and equatorial radii of the horizon.
Thus rough estimates may still be possible 
even for appreciable angular momentum.

The basic idea of a quasi-spherical approximation 
is to make a 2+2 decomposition of the space-time 
and linearize only those parts of the extrinsic curvature 
which vanish in spherical symmetry, cf.\ Bishop et al.\cite{BGLW}.
Thus when the linearized fields vanish, spherical symmetry is recovered in full.
This can be a highly dynamical situation; 
there will be no assumption of quasi-stationarity.
Likewise, there will be no assumption of an exactly spherical background.
Unlike previous work on null-temporal formulations\cite{BBH2,BGLW,BGLMW,W},
a dual-null formulation is adopted here,
i.e.\ a decomposition of the space-time 
by two intersecting foliations of null hypersurfaces.
This is adapted to the radiation problem in that 
the imposition of no ingoing radiation 
and the extraction of the outgoing radiation are immediate.
It also allows a remarkable simplification 
from partial to ordinary differential equations.

A general Hamiltonian theory of dual-null dynamics\cite{dn} 
has been applied to Einstein gravity\cite{dne} and is summarized as follows.
Denoting the space-time metric by $g$
and labelling the null hypersurfaces by $x^\pm$, 
the normal 1-forms $n^\pm=-dx^\pm$ therefore satisfy
\begin{equation}
g^{-1}(n^\pm,n^\pm)=0.
\end{equation}
The relative normalization of the null normals 
may be encoded in a function $f$ defined by
\begin{equation}
e^f=-g^{-1}(n^+,n^-).
\end{equation}
Then the induced metric on the transverse surfaces,
the spatial surfaces of intersection, is found to be
\begin{equation}
h=g+2e^{-f}n^+\otimes n^-
\end{equation}
where $\otimes$ denotes the symmetric tensor product.
The dynamics is described by Lie transport 
along two commuting evolution vectors $u_\pm$:
\begin{equation}
[u_+,u_-]=0.
\end{equation}
Specifically, the evolution derivatives, 
to be discretized in a numerical code, are
\begin{equation}
\Delta_\pm=\bot L_{u_{\pm}}
\end{equation}
where $\bot$ indicates projection by $h$ and $L$ denotes the Lie derivative.
There are two shift vectors 
\begin{equation}
s_\pm=\bot u_\pm.
\end{equation}
In a coordinate basis $(u_+,u_-;e_i)$ such that
$u_\pm=\partial/\partial x^\pm$, 
where $e_i=\partial/\partial x^i$ is a basis for the transverse surfaces,
the metric takes the form
\begin{eqnarray}
g&=&h_{ij}(dx^i+s_+^idx^++s_-^idx^-)\otimes\nonumber\\
&&(dx^j+s_+^jdx^++s_-^jdx^-)
-2e^{-f}dx^+\otimes dx^-.
\end{eqnarray}
Then $(h,f,s_\pm)$ are configuration fields
and the independent momentum fields are found to be linear combinations of
\begin{eqnarray}
\theta_\pm&=&*L_\pm{*}1\\
\sigma_\pm&=&\bot L_\pm h-\theta_\pm h\\
\nu_\pm&=&L_\pm f\\
\omega&=&\textstyle{1\over2}e^fh([l_-,l_+])
\end{eqnarray}
where $*$ is the Hodge operator of $h$ 
and $L_\pm$ is shorthand for the Lie derivative along the null normal vectors 
\begin{equation}
l_\pm=u_\pm-s_\pm=e^{-f}g^{-1}(n^\mp).
\end{equation}
Then the functions $\theta_\pm$ are the expansions,
the traceless bilinear forms $\sigma_\pm$ are the shears,
the 1-form $\omega$ is the twist,
measuring the lack of integrability of the normal space,
and the functions $\nu_\pm$ are the inaffinities, 
measuring the failure of the null normals to be affine. 
The fields $(\theta_\pm,\sigma_\pm,\nu_\pm,\omega)$ 
encode the extrinsic curvature of the dual-null foliation.
These extrinsic fields are unique up to duality $\pm\mapsto\mp$
and diffeomorphisms which relabel the null hypersurfaces, i.e.\
$dx^\pm\mapsto e^{\lambda_\pm}dx^\pm$
for functions $\lambda_\pm(x^\pm)$.

It is also useful to decompose $h$ into 
a conformal factor $r$ and a conformal metric $k$ by
\begin{equation}
h=r^2k
\end{equation}
such that 
\begin{equation}
\Delta_\pm\hat{*}1=0
\end{equation}
where $\hat{*}$ is the Hodge operator of $k$, satisfying ${*}1=\hat{*}r^2$.
Denoting the covariant derivative of $h$ by $D$,
the Ricci scalar of $h$ is found to be
\begin{equation}
R=2r^{-2}(1-D^2\ln r)
\end{equation}
by using the coordinate freedom on a given surface 
to fix $k$ as the metric of a unit sphere.

The dual-null Hamilton equations and integrability conditions 
for vacuum Einstein gravity have been given previously\cite{dne} 
in a slightly different notation, so will not be repeated here.
They are linear combinations of the vacuum Einstein equation
and a first integral of the contracted Bianchi identity. 
This is the vacuum Einstein system in first-order dual-null form.
The vacuum case suffices for the application, outside the black holes.

In spherical symmetry, $(s_\pm,\sigma_\pm,\omega,D)$ vanish,
while $(h,f,\theta_\pm,\nu_\pm,\Delta_\pm)$ are generally non-zero, 
e.g.\cite{sph,1st}.
The quasi-spherical approximation will therefore consist of linearizing in 
$(s_\pm,\sigma_\pm,\omega,D)$. 
In practice, one truncates the equations 
by setting to zero any second-order terms in $(s_\pm,\sigma_\pm,\omega,D)$.
This greatly simplifies the equations, leaving the momentum definitions as
\begin{eqnarray}
\Delta_\pm r&=&\textstyle{1\over2}r\theta_\pm\\
\Delta_\pm f&=&\nu_\pm\\
\Delta_\pm k&=&r^{-2}\sigma_\pm\\
\Delta_+s_--\Delta_-s_+&=&2e^{-f}h^{-1}(\omega) 
\end{eqnarray}
and the remaining equations as
\begin{eqnarray}
\Delta_\pm\theta_\pm&=&-\nu_\pm\theta_\pm-\textstyle{1\over2}\theta_\pm^2\\
\Delta_\pm\theta_\mp&=&-\theta_+\theta_--e^{-f}r^{-2}\\
\Delta_\pm\nu_\mp&=&-\textstyle{1\over2}\theta_+\theta_--e^{-f}r^{-2}\\
\Delta_\pm\sigma_\mp&=&
\textstyle{1\over2}(\theta_\pm\sigma_\mp-\theta_\mp\sigma_\pm)\\
\Delta_\pm\omega&=&-\theta_\pm\omega
\pm\textstyle{1\over2}(D\nu_\pm-D\theta_\pm-\theta_\pm Df).
\end{eqnarray}
This follows immediately from the full equations\cite{dne},
the above expression for $R$ and the fact that 
\begin{equation}
\Delta_\pm=\bot L_\pm 
\end{equation}
in this truncation.
One may take quasi-spherical coordinates $x^i=(\vartheta,\varphi)$
on the transverse surfaces such that 
$\hat{*}1=\sin\vartheta d\vartheta\wedge d\varphi$,
the standard area form of a unit sphere.
Then $r$ is the quasi-spherical radius.

The shear equations, composed into a second-order equation for $k$, become
\begin{equation}
\Box k=0
\end{equation}
where $\Box$ is the quasi-spherical wave operator:
\begin{equation}
\Box\phi=-2e^f\left(\Delta_{(+}\Delta_{-)}\phi
+2r^{-1}\Delta_{(+}r\Delta_{-)}\phi\right).
\end{equation}
Thus the conformal metric $k$ satisfies the quasi-spherical wave equation.
Then $k$ may be interpreted as encoding the gravitational radiation. 
In particular, fixing $u_+$ to be the outgoing direction, 
the Bondi news at future null infinity $\Im^+$ is 
essentially $r^{-1}\sigma_-$\cite{mono}, as described explicitly below.
Likewise, the no-ingoing-radiation condition is just $r^{-1}\sigma_+=0$ 
at past null infinity $\Im^-$.
That $k$ generally encodes the free gravitational data 
was suggested by d'Inverno \& Stachel\cite{dS}
and has been rediscovered by various authors, e.g.\cite{W,dne}.

The dual-null initial-data formulation is based on a spatial surface $\Sigma$
and the null hypersurfaces $\Sigma_\pm$ locally generated from $\Sigma$ 
in the $u_\pm$ directions.
The structure of the field equations shows that one may specify 
$(h,f,\theta_\pm,\omega)$ on $\Sigma$,
$(\sigma_+,\nu_+)$ on $\Sigma_+$,
$(s_-,\sigma_-,\nu_-)$ on $\Sigma_-$
and $s_+$ in $U$, a region to the future of $\Sigma_\pm$.
In particular, the initial data is freely specifiable.
There are no constraints as in the 3+1 formulation;
these have been converted into evolution equations along $\Sigma_\pm$,
which even in the general (non-quasi-spherical) case 
can be solved in closed form\cite{sol}.

\begin{figure}
\centerline{\epsfxsize=5cm \epsfbox{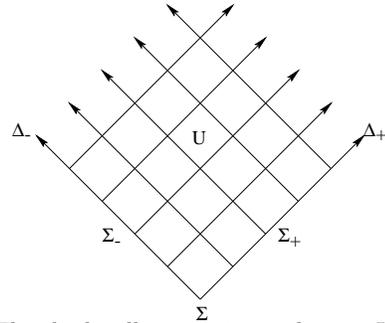}}
\caption{The dual-null integration scheme.
Initial data is prescribed on a spatial surface $\Sigma$ 
and the null hypersurfaces $\Sigma_+$ and $\Sigma_-$ generated from it.}
\label{id}
\end{figure}

A numerical integration scheme runs as follows,
as depicted in Fig.\ref{id}.
First integrate the $\Delta_+$ equations from $\Sigma$ 
to obtain the full data on $\Sigma_+$.
Then integrate the $\Delta_-$ equations one step 
along each ingoing null hypersurface, 
generating a new null hypersurface $\Sigma_+'$.
Then repeat:  integrate the $\Delta_+$ equations along $\Sigma_+'$ 
to obtain the full data on $\Sigma_+'$, and so on.
In practice, some interpolation between the two integrations is useful.
There are many ways to perform the integrations in a different order,
allowing flexibility which can be used, for instance, to avoid singularities.
Any such scheme gives two estimates of $(r,k,f,\theta_\pm,\omega)$ 
at each point,
since some of the equations play the role of integrability conditions.
Thus one could ignore such equations to obtain a 
free rather than constrained integration scheme.
This allows numerous internal checks on the accuracy of the numerical code,
analogous to those of 3+1 integration schemes, e.g.\ Choptuik\cite{C}.

The equations for $\Delta_\pm(r,f,\theta_\pm,\nu_\pm)$, 
the quasi-spherical equations,
decouple from the remaining equations.
Thus there is a quasi-spherical background 
which may be found by integrating the quasi-spherical initial data. 
Since this background is independent of the linearized part,
one may economize when computing different evolutions on the same background.
It should be stressed that 
the quasi-spherical background is neither fixed in advance
nor necessarily spherical, e.g.\ $Dr\not=0$ in general.

To compute the outgoing radiation,
one now needs only to integrate the equations for 
$(\Delta_\pm k,\Delta_\pm\sigma_\mp)$,
i.e.\ the quasi-spherical wave equation for $k$.
It is remarkable that 
this entire integration scheme involves only ordinary differential equations.
The equations for $\Delta_\pm\omega$ are partial differential equations, 
containing transverse $D$ derivatives,
but the other equations decouple from 
the equations for $(s_\pm,\omega)$,
which therefore need not be solved for the radiation problem.
In short, 
most of the complexity of the system has been isolated and sidestepped.

\begin{figure}
\centerline{\epsfxsize=8cm \epsfbox{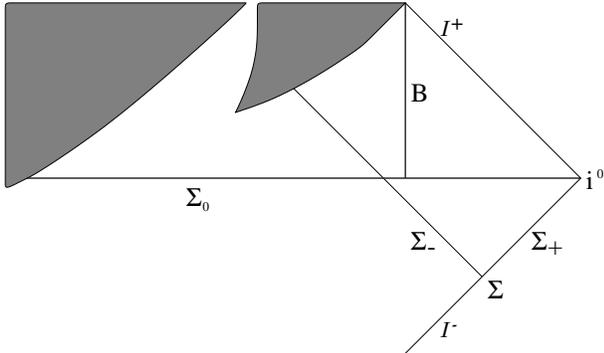}}
\caption{Application to coalescing black holes.
Shading indicates regions containing trapped surfaces,
with the outermost trapped region being that of the coalesced black hole
and the inner trapped region that of one of the original black holes.
Initial data on the ingoing null hypersurface $\Sigma_-$
may be extracted from a conventional 3+1 code
based on an initial spatial hypersurface $\Sigma_0$ 
with outer boundary $B$.}
\label{bh}
\end{figure}

Moreover, one may use a conformal transformation 
to obtain a scheme which is more accurate at large distances.
Using the conformal factor 
\begin{equation}
\Omega=r^{-1}
\end{equation}
the rescaled expansions and shears 
\begin{eqnarray}
\vartheta_\pm&=&r\theta_\pm\\
\varsigma_\pm&=&r^{-1}\sigma_\pm
\end{eqnarray}
are finite and generally non-zero at $\Im^\mp$ 
for an asymptotically flat space-time.
Rewriting the relevant equations yields the quasi-spherical equations
\begin{eqnarray}
\Delta_\pm\Omega&=&-\textstyle{1\over2}\Omega^2\vartheta_\pm\\
\Delta_\pm f&=&\nu_\pm\\
\Delta_\pm\vartheta_\pm&=&-\nu_\pm\vartheta_\pm\\
\Delta_\pm\vartheta_\mp
&=&-\Omega(\textstyle{1\over2}\vartheta_+\vartheta_-+e^{-f})\\
\Delta_\pm\nu_\mp
&=&-\Omega^2(\textstyle{1\over2}\vartheta_+\vartheta_-+e^{-f})
\end{eqnarray}
and the linearized equations
\begin{eqnarray}
\Delta_\pm k&=&\Omega\varsigma_\pm\\
\Delta_\pm\varsigma_\mp
&=&-\textstyle{1\over2}\Omega\vartheta_\mp\varsigma_\pm.
\end{eqnarray}
One may take $\Sigma_+$ to be either part of $\Im^-$,
as depicted in Fig.\ref{bh},
or at sufficiently large distance for numerical purposes.
Here, large distance means small $\Omega$.
For the quasi-spherical approximation to be valid at large distance,
one may fix 
$(f,\vartheta_\pm,k)=(0,\pm\sqrt{2},\epsilon)$ on $\Sigma$,
where $\epsilon
=d\vartheta\otimes d\vartheta+\sin^2\vartheta d\varphi\otimes d\varphi$ 
is the standard metric of a unit sphere,
and $\nu_+=0$ on $\Sigma_+$.
The remaining coordinate data is given by 
$\nu_-$ on the ingoing null hypersurface $\Sigma_-$, 
which is left free so that one may adapt the foliation of $\Sigma_-$ 
to the surfaces which are most spherical.

The no-ingoing-radiation condition is $\varsigma_+=0$ at $\Im^-$,
leaving the gravitational initial data as $\varsigma_-$ on $\Sigma_-$.
The outgoing radiation is found by computing $\varsigma_-$ at $\Im^+$,
which is essentially the Bondi news.
More precisely, the Bondi energy flux at $\Im^\mp$ would be\cite{mono}
\begin{equation}
\psi_\pm=-{e^f\vartheta_\mp||\varsigma_\pm||^2\over{64\pi}}
\end{equation}
where $||\sigma||^2=k^{ab}k^{cd}\sigma_{ac}\sigma_{bd}$
and such second-order terms are no longer being ignored.
That is, the energy supply would be 
\begin{equation}
\Delta_\pm E=\oint\hat{*}\psi_\pm
\end{equation}
where $E$ is the Bondi energy.
In summary, the outgoing waveforms and their energy may be computed 
by integrating 9 first-order ordinary differential equations and their duals,
or a subset in the case of free evolution.
For numerical purposes, this is a dramatic simplification.
Numerical implementation of this scheme is in progress\cite{SH}.

Before concluding, it should be noted that the domain of validity 
of the quasi-spherical approximation is not known in a precise sense.
The guarantee is simply that spherically symmetric Einstein gravity 
is recovered in full when the linearized fields vanish.
For the usual perturbative approximations,
one may check successive orders of approximation to compare accuracy,
but for the quasi-spherical approximation, 
the corresponding second-order approximation would be full Einstein gravity.
If one wishes to know whether a given space-time is sufficiently spherical,
the rough answer is that there should be a 2+2 decomposition such that 
the fields to be linearized are small compared to the remaining part.
This depends on the choice of transverse surfaces,
so that there will be some art to choosing the 2+2 foliation 
for optimal accuracy.
For the Kerr black hole, the quasi-spherical null coordinates 
of Pretorius \& Israel\cite{PI} may be useful.
For a coalesced black hole, one might base the foliation 
on the marginal surfaces which locally define it, 
i.e.\ use the coordinate freedom in $\nu_-$ on $\Sigma_-$
so that the foliation contains a marginal surface.
There are some general laws of black-hole dynamics 
in terms of marginal surfaces 
and the trapping horizons they generate\cite{bhd,MH},
including that outer trapping horizons are achronal
and therefore cannot causally influence $\Im^+$.
Thus $\Sigma_-$ need not extend inside the trapped region 
in order for the domain $U$ of integration to reach all of $\Im^+$.

To conclude, the intended scenario is an asymptotically flat space-time 
containing coalescing black holes,
with an ingoing null hypersurface $\Sigma_-$ 
chosen to intersect the coalesced black hole,
i.e.\ the region of future trapped surfaces enclosing the original black holes.
The initial data on $\Sigma_-$ may be determined by 
extracting the relevant data from a conventional 3+1 numerical computation
from an initial spatial hypersurface $\Sigma_0$,
smoothed off to the past of $\Sigma_0$.
The smoothing may be expected not to affect the results significantly 
if $\Sigma_-\cap\Sigma_0$ is sufficiently outside the black holes;
there will be some spurious radiation at $\Im^+$ at early times,
but not at the relevant late times, 
as this would involve backscatter of backscatter.

The advantage of this procedure is that 
it avoids the outer-boundary problems which plague the conventional codes.
As depicted in Fig.\ref{bh}, 
the outer boundary $B$ cannot causally influence $\Sigma_-$
if it intersects $\Sigma_0$ inside $B$.
Thus the scheme requires only clean data from the 3+1 computation,
uncontaminated by outer-boundary problems.
A code implementing the scheme may be regarded as a black box which, 
taking input from any other code 
from which the required data on $\Sigma_-$ can be extracted,
computes approximate waveforms for the gravitational radiation.

This suggests a quite general proposal 
to compute outgoing gravitational radiation 
from a 3+1 computation by a conformal dual-null code 
which extracts data on an ingoing null hypersurface 
intersecting the initial spatial hypersurface inside its outer boundary.
One might expect this to be simpler than 
the usual matching on a temporal hypersurface\cite{BBH2,BGLW,BGLMW,W}
since the outer boundary is avoided 
and the problem is merely of extraction rather than dynamic matching.
At present, there seems to be neither a general conformal dual-null code
nor work on data extraction on a null hypersurface,
though the null-temporal formulation can presumably be adapted\cite{W}.
These are tractable projects which would allow 
accurate computation of gravitational waveforms from coalescing black holes.

\bigskip\noindent
Acknowledgements.
The author thanks Pablo Laguna, Luis Lehner, Keith Lockitch, 
Hisa-aki Shinkai and Jeff Winicour for discussions,
and Abhay Ashtekar and the Center for Gravitational Physics and Geometry 
for hospitality.
Research supported by the National Science Foundation under award PHY-9800973.

\end{document}